\documentclass[letter]{jpsj4}

\title{Microwave Surface Impedance Measurements of LiFeAs Single Crystals}

\author{Yoshinori Imai\thanks{E-mail address: imai@maeda1.c.u-tokyo.ac.jp}$^{1,4}$, Hideyuki Takahashi$^{1,4}$, Kentaro Kitagawa$^{2,4}$, Kazuyuki Matsubayashi$^{2,4}$, Noriyuki Nakai$^{3,5}$, Yuki Nagai$^{3,4,5}$, Yoshiya Uwatoko$^{2,4}$, Masahiko Machida$^{3,4,5}$, Atsutaka Maeda$^{1,4}$
}

\inst{$^{1}$ Department of Basic Science, the University of Tokyo, 3-8-1 Komaba, Meguro, Tokyo 153-8902, Japan\\
$^{2}$ Institute for Solid State Physics, University of Tokyo, 5-1-5 Kashiwanoha, Kashiwa, Chiba 277-8581, Japan\\
$^{3}$ CCSE, Japan Atomic Energy Agency, Taito, Tokyo 110-0015, Japan\\
$^{4}$ Transformative Research-Project on Iron Pnictides (TRIP), Japan Science and Technology Agency, 5 Sanbancho, Chiyoda, Tokyo 102-0075, Japan\\
$^{5}$  CREST (JST), Kawaguchi, Saitama 332-0012, Japan
}

\abst{We report results of microwave surface impedance measurements of LiFeAs single crystals.  The in-plane penetration depth depends on temperature exponentially at low temperatures, which strongly suggests that this material has the nodeless superconducting gap.  The temperature dependence of the superfluid density indicates that LiFeAs is a multi-gap superconductor with at least two isotropic gaps.  In addtion, the real part of the microwave conductivity exhibits a large enhancement below $T_\mathrm{c}$, indicating that the quasi-particle relaxation time, $\tau$, increases rapidly below $T_\mathrm{c}$.  We believe that this enhancement is rather common to all superconductors where an inelastic scattering is dominant above $T_\mathrm{c}$, irrespective of the strength of the electron correlation.}

\kword{microwave surface impedance, penetration depth, superconducting gap structure, iron-based superconductors, LiFeAs}

\begin{document}
\maketitle



The discovery of a new superconducting material family, iron-based superconductors, has generated a great deal of interest.\cite{Kamihara08}.  The highest superconducting transition temperature, $T_\mathrm{c}$, in this system reached 55 K in SmFeAsO in a subsequent study.\cite{Ren08}  Up to now, several structural types of iron-based superconductors have been discovered.\cite{RotterPRL08, Tapp08, Wu08, Ogino09}
LiFeAs is one of the few iron based superconductors that show superconductivity without doping  additional carriers.\cite{Tapp08}  As are the parent compounds in  $Ln$FeAsO$_{1-x}$F$_x$ (``1111" system) or Ba$_{1-x}$K$_x$Fe$_2$As$_2$ (``122" system), LiFeAs consists of (Fe$_2$As$_2$)$^{2-}$  two dimensional layers, and these three materials have an isoelectronic state.  In addition, similar electronic structures and magnetic ground states are obtained in these three materials by the band calculations.\cite{Singh08, Kuroki08, SinghPRB08}  Nonetheless, the superconductivity does not appear in the parent compounds of 1111 and 122, but appears in LiFeAs.  Additionally, there is no experimentally-indicated evidence for the magnetic transition, which corresponds to the spin-density wave order in the parent compounds of 1111 and 122,\cite{Cruz08}  in LiFeAs,\cite{Tapp08, PrattPRB09} contrary to the expectation from the band calculation.\cite{SinghPRB08} 
It is of a great interest why the physical properties in these materials are different in spite of the similarity in the crystal and electronic structures, and whether the structure of the superconducting order parameters is the same in these superconductors.

In general, to identify the structure of superconducting order parameters, the presence or absence of nodes in the gap function has a great importance.  The sign-reversing $s_{\pm}$ state, where the nesting between hole and electron bands plays an important role, has been proposed as  the most reasonable superconducting pairing in these materials, based on the experimental results in 1111 and 122.\cite{Kondo08, DingEPL08, Hashimoto091111, Hashimoto09122,Nakai08, Yashima09}.
In LiFeAs, however, the reported experimental results are controversial: Inosov $et \  al.$ reported LiFeAs has a single isotropic superconducting gap based on angle-resolved photoemission (ARPES) and small-angle neutron scattering (SANS) measurements.\cite{Inosov10}  On the other hand, some papers reported that there were multi superconducting gaps in LiFeAs. \cite{Borisenko10, WeiPRB10}
In this Letter, we present surface-impedance measurements in LiFeAs single crystals, and discuss the structure of the superconducting gap based on the in-plane penetration depth and the relaxation time of quasiparticle below $T_\mathrm{c}$.
In this study, we grew the crystals of LiFeAs by a self-flux method starting with 2:1:2 of elements. The materials were put into BN crucible and sealed in a doubled-wall quartz tube in argon atmosphere. These were heated once to 1000$^\circ$C and crystallized by gradually cooled down from 930 to 400$^\circ$C in 80 hours and to room temperature in 1 day.  Scanning tunnel microscopy proves the stoichiometry of the cleaved surface within an error of 1\% \cite{HanaguriXX}.  
All handlings have been performed under purified argon except after the sample was wrapped by grease in order to place the sample on the stage.  This crystal shows superconductivity at 19.0 K (onset; $T_\mathrm{c}^\mathrm{onset}$) and 17.9 K (zero resistivity; $T_\mathrm{c}^\mathrm{zero}$), and has a resistance ratio, $r=R(300\mathrm{K})/R(20\mathrm{K})$, as large as 45, which indicates the high quality of the single crystal.  
The surface impedance, $Z_\mathrm{s}$, was measured by a cavity perturbation method with a hot-finger technique.\cite{MaedaReview05}  Three single crystals of LiFeAs are used in our measurements, and the properties are summarized in Table \ref{Samplespec}.  We used a 19 GHz TE$_{011}$-mode O-free copper cavity with a quality factor, $Q \approx 60000$.  To measure the surface impedance of a small single crystal with high precision, the cavity resonator is soaked in the superfluid $^4$He at 1.5 K.  The LiFeAs single crystal is placed in the antinode of the microwave magnetic field, $H_ \omega$(// $c$-axis), so that the shielding current, $I_\omega$, is excited in the $ab$ planes [see schematic figure in Fig. 1].  The inverse of quality factor, 1/$Q$, and the shift in the resonance frequency, $\Delta f \equiv f_{s0} - f_{0}$ (where $f_{s0}$ and $f_{0}$ are resonance frequencies with and without a sample, respectively), are proportional to the surface resistance, $R_\mathrm{s}$, and the change in the surface reactance, $X_\mathrm{s}$, respectively.  
In our frequency range of $\omega/2\pi \approx 19$ GHz, the conductivity, $\sigma=\sigma_1 + i\sigma_2$, can be extracted from $Z_s$ through the relation that is valid for the so-called skin depth regime, where the skin depth,  $\delta$, is much shorter than the sample size:
\begin{equation}
Z_s = R_s - i X_s = \sqrt{\frac{-i \mu_0 \omega}{\sigma_1 + i \sigma_2}}.
\label{eqzs}
\end{equation}
In the superconducting state, the surface reactance is a direct measure of the superﬂuid density, $n_s$, via $X_s$($T$)=$\mu_0 \omega \lambda_{ab}$($T$) and $\lambda_{ab}^{-2}(T)=\mu_0 n_s(T)e^2/m^*$.  In the normal state,  Eq. (\ref{eqzs}) gives $R_s(T)=X_s(T)=(\mu_0\omega/2\sigma_1)^{1/2}$ since $\sigma_1 = ne^2\tau/m^* \gg \sigma_2$, where $n$ is the total density of carriers with effective mass $m^*$.  This relation can be used to determine $X_s(0)/X_s(T_\mathrm{c})$, which allows us to determine $\lambda_{ab}(T)/\lambda_{ab}(0)$ without any assumptions\cite{Shibauchi94}. 
In the simple two-fluid model, the real part of the conductivity, $\sigma_1$, is determined by the quasi-particle dynamics.
The quasiparticle scattering time, $\tau$, can be estimated from $\sigma_1$ and $\sigma_2$ through the relation: 
\begin{equation}
\frac{1}{1-f_s}(\omega\tau)^2 -\frac{\sigma_2}{\sigma_1}(\omega\tau)+\frac{f_s}{1-f_s}=0  ,
\label{eqtau}
\end{equation}
where $f_s \equiv n_s/n$ is the superconducting fraction.

\begin{figure}
\begin{center}
\includegraphics{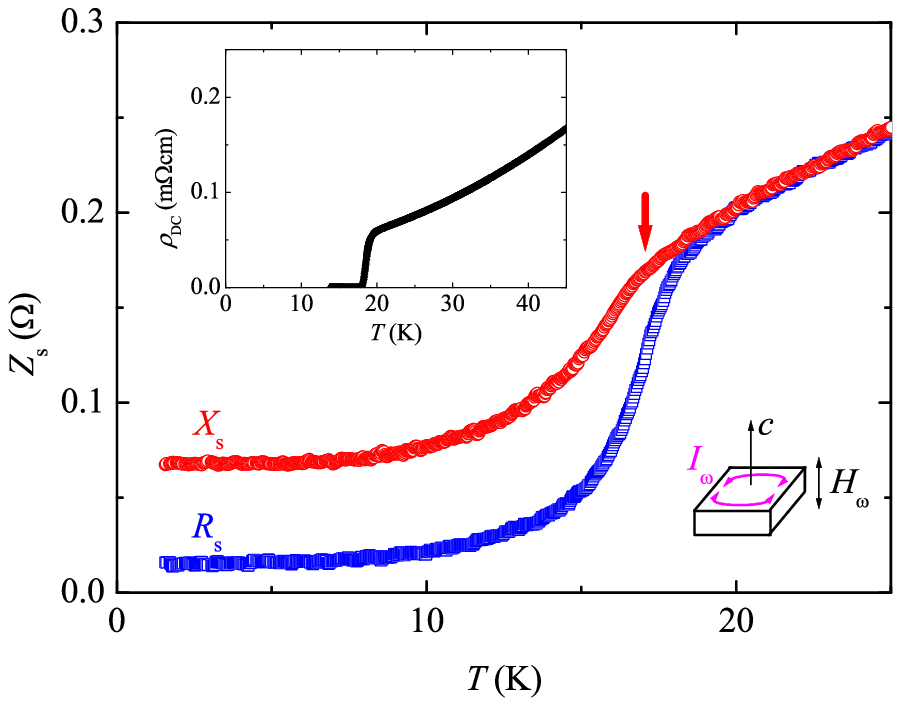}
\end{center}
\caption{ }
\label{zs}
\end{figure}

\begin{figure}
\begin{center}
\includegraphics[width=.9\linewidth]{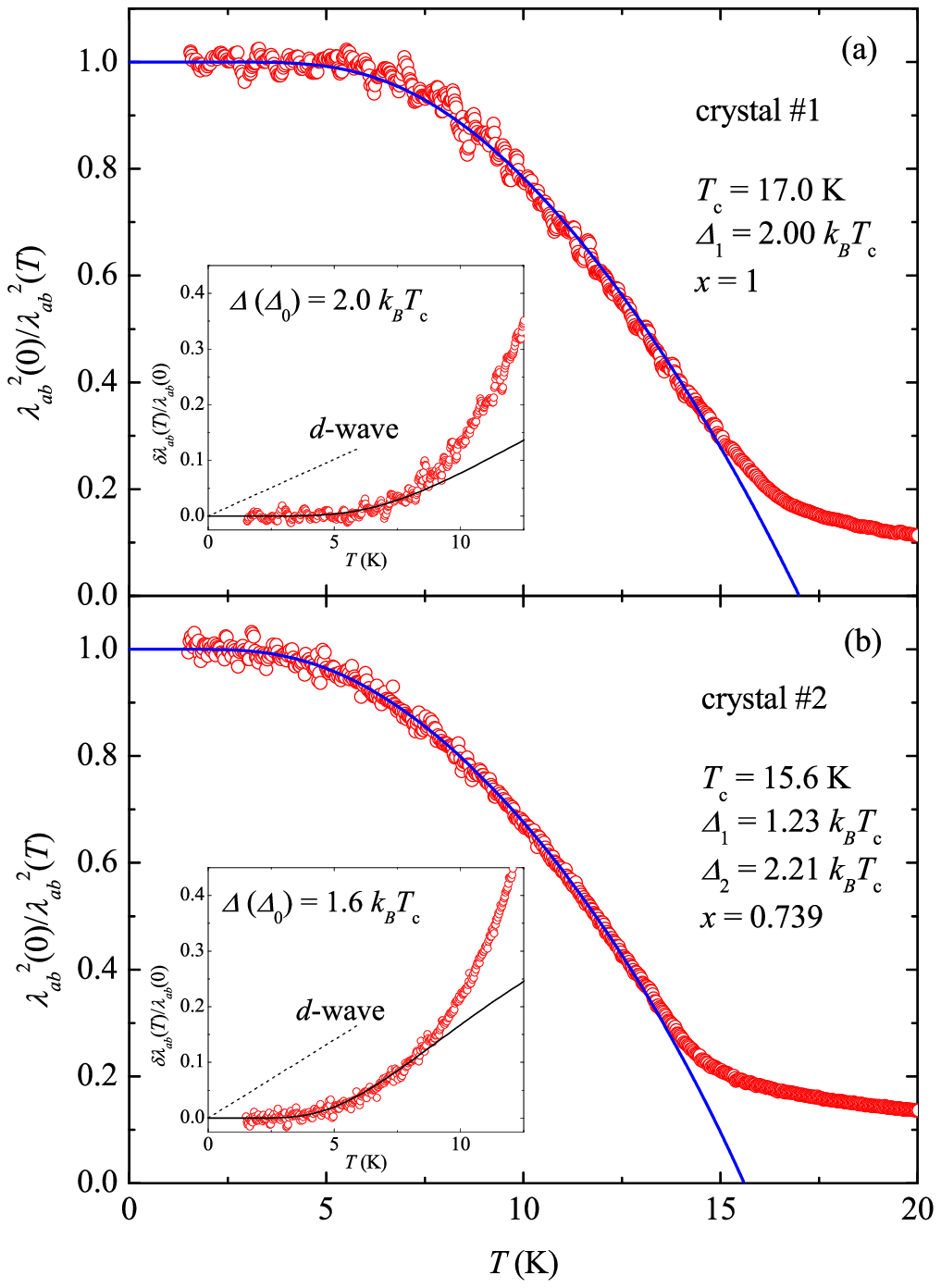}
\end{center}
\caption{ }
\label{ns}
\end{figure}

Figure \ref{zs} shows the typical temperature dependence of  $Z_\mathrm{s}$ for crystal \#1.  The superconducting transition temperature, $T_\mathrm{c}$, is defined as the temperature where $X_s$ starts to deviate from the normal state behavior.  $T_\mathrm{c}$ of crystal \#1 is represented by the large red downward arrow in Fig. \ref{zs}, and estimated values are summarized in Table \ref {Samplespec}.  As expected from Eq. (\ref{eqzs}), the temperature dependencies of the real and the imaginary parts above $T_\mathrm{c}$ are identical.  
In the superconducting state, $\lambda_{ab}$ is obtained from the surface reactance, via $X_s$($T$)=$\mu_0 \omega \lambda_{ab}$($T$).  $\lambda_{ab} (0)$ of crystals \#1-3 are estimated at the values in Table \ref{Samplespec}.  These values are slightly large compared to the other experiments.\cite{PrattPRB09, Inosov10} 
The normalized change in the in-plane penetration depth, $\delta\lambda_{ab}(T) = \lambda_{ab}(T)-\lambda_{ab}(0)$, of crystals \#1 and \#2 at low temperatures is shown in the insets of Fig. \ref{ns}.  It is clear that $\delta\lambda_{ab}(T)$ has a flat dependence at low temperatures.  First, we compare these data with the expectations in unconventional superconductors with nodes in the superconducting gap.  In clean superconductors with line nodes, thermally excited quasiparticles near the gap nodes give rise to the $T$-linear temperature dependence of $\delta\lambda_{ab}(T)$ at low temperatures, as was observed in YBa$_2$Cu$_3$O$_{7-\delta}$ single crystals with $d$-wave symmetry\cite{Hardy93}.  In the case of a superconductor with $d$-wave symmetry, the penetration depth satisfies the relation of $\delta\lambda_{ab}(T)/\lambda_{ab}(0) \approx (\ln 2 / \Delta_0)k_\mathrm{B} T$ \cite{MaedaReview05} where $\Delta_0$ is the maximum of the energy gap $\Delta(\mathbf{k})$.  This linear temperature dependences with a gap ratio of $\Delta_0/k_\mathrm{B}T_\mathrm{c} \approx$ 2.0 (crystal \#1), 1.6 (crystal \#2), which are shown by the dashed line in the insets of Fig. \ref{ns}, clearly deviate from the data of the LiFeAs single crystals.  In superconductors without any nodes, the quasiparticle excitation is a thermally-activated type, which gives an exponential dependence
\begin{equation}
\frac{\lambda_{ab}(T)}{\lambda_{ab}(0)} \approx \sqrt{\frac{\pi \Delta}{2k_\mathrm{B} T}}\exp (-\frac{\Delta}{k_\mathrm{B} T})
\label{eqlambda}
\end{equation}
for $T \le T_\mathrm{c}/2$\cite{Halbritter71}.  
Fitting of the experimental data to Eq. (\ref{eqlambda}), which is shown by the solid line in the insets of Fig. \ref{ns}, enables us to estimate the minimum energy $\Delta_\mathrm{min}$ required for quasiparticle excitations at $T=0$ K, $i. e.$, $2\Delta_\mathrm{min}/k_\mathrm{B}T_\mathrm{c}=$ 4.0 (crystal \#1), 3.2 (crystal \#2).

The normalized superfluid density $n_s/n=\lambda_{ab}^2(0)/\lambda_{ab}^2(T)$ of crystals \#1 and \#2 is also plotted as a function of temperature in the main panels of Figs. \ref{ns} (a) and (b), respectively.  As is the temperature dependence of $\delta \lambda_{ab}$, the low-temperature behavior is flat, indicating a nodeless superconducting gap.  Since the iron-based superconductors have the multiband electronic structure\cite{Singh08}, we try to fit the whole temperature dependence of $n_s$ with a simple two-gap model $n_s(T)=xn_{s1}(T)+(1-x)n_{s2}(T)$\cite{KoganPRB04, Fletcher05, Hashimoto091111, Hashimoto09122}.  The band 1 (2) has the superfluid density, $n_{s1}$ ($n_{s2}$), which is determined by the gap, $\Delta_1$ ($\Delta_2$), and $x$ defines the relative weight of each band to $n_s$.  The best fitted results are summarized in Table \ref{Samplespec}.  Except for crystal \#1, the temperature dependence of the superfluid density cannot be wholly represented by a single-gap calculation.   This suggests that LiFeAs is a superconductor with at least two nodeless superconducting gaps.  It should be noted that the values of  $\Delta_1$ and $\Delta_2$ in crystals \#2 and \#3 are comparable to ones reported by ARPES.\cite{Borisenko10}  According to the comparison with the result of ARPES,\cite{Borisenko10} $\Delta_1$ and $\Delta_2$ correspond to the superconducting gaps around hole-like and electron-like Fermi surfaces, respectively.  On the other hand, one ``large" superconducting gap is observed in crystal \#1 unlike the other two crystals.  
It should be noted that $\Delta_1$ is almost the same among all three crystals investigated,  whereas $\Delta_2$ is different between crystals \#2 and \#3.  Therefore, we interpret that $\Delta_2$ becomes almost equal to $\Delta_1$ in crystal \#1, leading to the apparent absence of the second gap, $\Delta_2$.  Indeed, the crystal with larger $\Delta_2$ has smaller $\lambda_{ab}$.
The difference of $\Delta_2$ may result from the delicate difference among some pieces of LiFeAs single crystals, possibly resulting from the differences of Li contents and/or the surface state, and so on.  After all, we conclude that LiFeAs is a two-gap superconductor as a general feature.

\begin{figure}
\begin{center}
\includegraphics{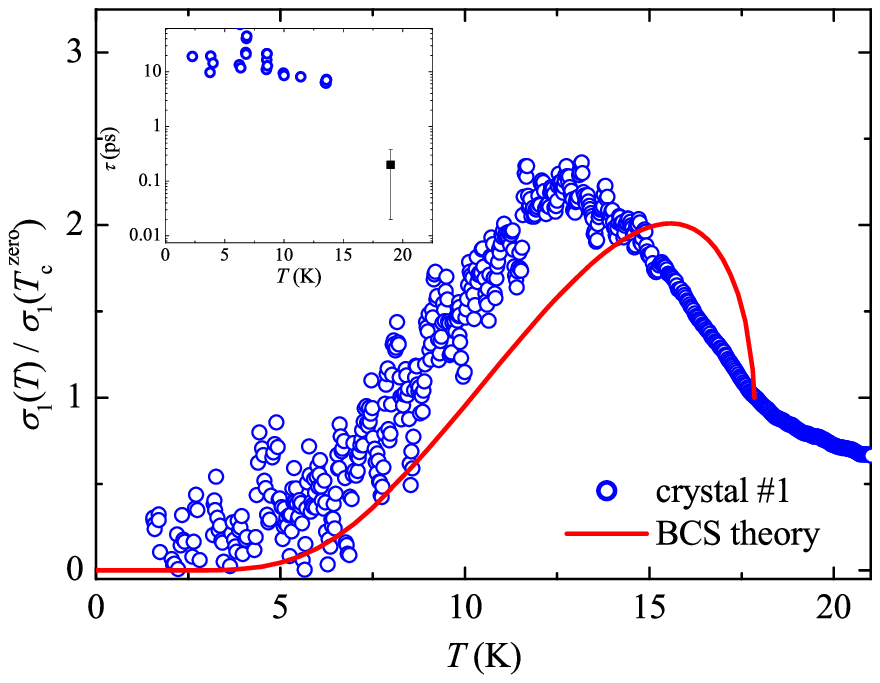}
\end{center}
\caption{ }
\label{sigma1}
\end{figure}

Figure \ref{sigma1} shows the temperature dependence of the normalized quasiparticle conductivity, $\sigma_1(T)/\sigma_1(T_\mathrm{c}^\mathrm{zero})$, of crystal \#1 which is extracted from $Z_s$ by Eq. (\ref{eqzs}).  $\sigma_1$ shows  a large enhancement below $T_\mathrm{c}$.  In conventional BCS superconductors, $\sigma_1$ shows an enhancement called as the ``coherence peak".\cite{KleinPRB94}  In comparison with the coherence peak [blue solid line in Fig. \ref{sigma1}]\cite{Zimmermann91}, this peak is not asymmetric, and considerably large and broad, and shifts slightly toward low temperatures.  
In addition, a peak near $T_\mathrm{c}$ do not observed in the NMR relaxation rate of the same material.\cite{Li111NMRJPSJ10, KitagawaXX}  Since the coherence peak becomes suppressed with increasing frequency in the framework of the BCS theory, the result of the NMR experiment strongly suggests that this enhancement has a different origin.  
The absence of the coherence peak in $\sigma_1(T)$ suggests that two superconducting gaps in LiFeAs have different signs($s_\pm$).


Similar enhancement is observed in $\sigma_1(\omega, T)$ in cuprate superconductors.  This is understood as the result of the competition between increasing the quasiparticle scattering time and the decreasing quasiparticle density, $n_n(T)=n-n_s(T)$, with decreasing temperature.  In this context, the quasiparticle scattering time, $\tau$, of crystal \#1 in the superconducting state, which is calculated by Eq. (\ref{eqtau}), is shown in the inset of Fig. \ref{sigma1}.  In the normal state, $\tau$ is  estimated at about $2\times 10^{-13}$ sec from $\rho_\mathrm{DC}(=\frac{m^{*}}{ne^2\tau})$, which is shown by the filled square in Fig. \ref{sigma1}.  With the superconducting transition, $\tau$ observes a sharp increase by 2 orders of magnitude.  
The rapid increase of $\tau$ below $T_\mathrm{c}$ suggests that a gap opens in the excitation spectrum of a dominant quasiparticle scatterer by the superconducting transition.  Thus, it indicates that the dominant inelastic scattering in the normal state of LiFeAs has an electronic origin.   
Besides cuprates, the similar peak is commonly observed in heavy-electron superconductors \cite{Ormeno02} and other iron-based superconductors (1111 and 122)\cite{Hashimoto091111, Hashimoto09122}.  All of these materials have strong electron correlation.
LiFeAs, on the other hand, is considered to be a fermi-liquid material with a relatively weak electron correlation because of the following reasons: First, the temperature dependence of $\rho_\mathrm{DC}$ is proportional to the square of temperature, which is a characteristic behavior in the Fermi-liquid material.  Next, a band calculation of LiFeAs also describes the experimentally-acquired electronic structure rather well,\cite{SinghPRB08, Inosov10, Borisenko10} indicating that the band picture is appropriate in this material.  Therefore, the rapid increase of $\tau$ below $T_\mathrm{c}$ observed in LiFeAs shows that this phenomena is not characteristic of strongly-correlated superconductors, but is rather common to superconductors where the inelastic scattering is dominant above $T_\mathrm{c}$, irrespective of the strength of the electron correlation.

In conclusion, we have measured the surface impedances in LiFeAs single crystals.  The temperature dependence of the penetration depth at low temperatures is the thermally-activated type, which shows that this material has a superconducting gap without any nodes.  The fitting of the temperature dependence of the superfluid density by the phenomenological model reveals that LiFeAs has, at least, two isotropic superconducting gaps.
In addition, the temperature dependence of $\sigma_1$ shows a broad peak below $T_\mathrm{c}$.  The origin of this peak is not the coherence peak, but the rapid increase of the quasiparticle scattering time below $T_\mathrm{c}$.    
The observation of the enhancement of $\tau$ in LiFeAs, which is considered to be a typical fermi-liquid material, indicates that this behavior is rather common to superconductors where the inelastic scattering is dominant above $T_\mathrm{c}$, irrespective of the strength of the electron correlation.

\appendix

We thank Mr. Joji Nasu, Prof. Yusuke Kato, and Prof. Masashi Takigawa for the fruitful discussions.

\newpage

Fig. \ref{zs}: (color online) Temperature dependence of $R_\mathrm{s}$ and $X_\mathrm{s}$ at 19 GHz in crystal \#1.  The large red downward arrow exhibits $T_\mathrm{c}$ of crystal \#1. The inset shows the temperature dependence of the dc resistivity.\\

Fig. \ref{ns}: (color online) Main panels: $ \lambda_{ab}^2$(0)/$\lambda_{ab}^2$($T$) for (a) crystal \#1 and (b) crystal \#2 fitted to the simple two-gap model with (a) $\Delta_1 = 2.00 k_\mathrm{B}T_\mathrm{c}$ or (b) $\Delta_1 = 1.23 k_\mathrm{B}T_\mathrm{c}, \Delta_2 = 2.21 k_\mathrm{B}T_\mathrm{c}$ (blue solid line).  Above $T_\mathrm{c}$, the normal-state skin-depth contribution gives a finite tail.  Insets: $\delta \lambda_{ab}$($T$)/$\lambda_{ab}$(0) at low temperatures for (a) crystal \#1 and (b) crystal \#2 fitted to Eq. (\ref{eqlambda}) with $\Delta=$ (a) $2.0 k_\mathrm{B}T_\mathrm{c}$ or (b) $1.6 k_\mathrm{B}T_\mathrm{c}$ (solid line).  The dashed line represents the $T$-linear dependence expected in clean $d$-wave superconductors \cite{MaedaReview05} with maximum gap $\Delta_0=$ (a) $2.0 k_\mathrm{B}T_\mathrm{c}$ or (c) $1.6 k_\mathrm{B}T_\mathrm{c}$.\\

Fig. \ref{sigma1}: (color online) Temperature dependence of normalized quasiparticle conductivity at 19 GHz for crystal \#1.  The red solid line is a BCS calculation\cite{Zimmermann91} with $\tau(T_\mathrm{c})=2.0 \times 10^{-13}$ sec.  The inset shows the temperature dependence of $\tau$.  $\tau$ in the superconducting state is calculated by Eq. (\ref{eqtau}).  $\tau$ in the normal state, on the other hand, is estimated at about $2 \times 10^{-13}$ sec from the data of $\rho_\mathrm{DC}$.

\newpage

\begin{table}
\caption{Specifications of LiFeAs single crystals we studied.  $T_\mathrm{c}$ is defined as the temperature where $X_s$ starts to deviate from the normal state behavior.  $\Delta_1$, $\Delta_2$ and $x$ are determined by the fitting to the temperature dependence of the superfluid density using the phenomenological two-gap model.  Details are described in the text.}
\label{Samplespec}
\begin{center}
\begin{tabular}{cccccccc}
\hline\hline
\multicolumn{1}{c}{Crystal} & \multicolumn{1}{c}{$T_\mathrm{c}$ (K)} & \multicolumn{1}{c}{$\lambda_{ab}$ (nm)} & \multicolumn{1}{c}{$\Delta_1$ (meV)} & \multicolumn{1}{c}{$\Delta_2$ (meV)} & \multicolumn{1}{c}{$\Delta_1/k_\mathrm{B}T_\mathrm{c}$} & \multicolumn{1}{c}{$\Delta_2/k_\mathrm{B}T_\mathrm{c}$} & \multicolumn{1}{c}{$x$}\\
\hline
\verb #1 & 17.0 & $450\pm90$ & 2.93 & - & 2.00 & - & 1\\
\verb #2 & 15.6 & $510\pm90$ & 2.97 & 1.65 & 2.21 & 1.23 & 0.739\\
\verb #3 & 16.3 & $600\pm90$ & 2.98 & 1.10 & 2.12 & 0.785 & 0.896\\
\hline\hline
\end{tabular}
\end{center}
\end{table}

\  \\

\newpage

\newpage

\newpage

\bibliographystyle{jpsj} 

\hyphenation{Post-Script Sprin-ger}


\end{document}